\documentclass[aps,prl,superscriptaddress,twocolumn,amsmath,amssymb]{revtex4}

\pdfoutput=1

\usepackage{color}
\usepackage{graphicx}
\usepackage{bm}
\usepackage{commath}
\usepackage{amsmath}
\usepackage{latexsym}
\usepackage{graphicx}
\usepackage{csquotes}
\begin{document}

\title{Multilayer black phosphorus as a versatile mid-infrared electro-optic material}

\author{Charles Lin}
\email{charleschihchin.lin@mail.utoronto.ca}
\affiliation{The Edward S. Rogers Department of Electrical and Computer Engineering, University of Toronto, 10 King's College Road, Toronto, ON M5S 3G4, Canada}
\author{Roberto Grassi}
\affiliation{Department of Electrical and Computer Engineering, University of Minnesota, 200 Union Street SE, 4-174 Keller Hall, Minneapolis, MN 55455-0170, United States}
\author{Tony Low}
\affiliation{Department of Electrical and Computer Engineering, University of Minnesota, 200 Union Street SE, 4-174 Keller Hall, Minneapolis, MN 55455-0170, United States}
\author{Amr S. Helmy}
\affiliation{The Edward S. Rogers Department of Electrical and Computer Engineering, University of Toronto, 10 King's College Road, Toronto, ON M5S 3G4, Canada}

\date{\today}
\begin{abstract}
We investigate the electro-optic properties of black phosphorus (BP) thin films for optical modulation in the mid-infrared frequencies. Our calculation indicates that an applied out-of-plane electric field may lead to red-, blue-, or bidirectional shift in BP's absorption edge. This is due to the interplay between the field-induced quantum-confined Franz-Keldysh effect and the Pauli-blocked Burstein-Moss shift. The relative contribution of the two electro-absorption mechanisms depends on doping range, operating wavelength, and BP film thickness. For proof-of concept, simple modulator configuration with BP overlaid over a silicon nanowire is studied. Simulation result shows that operating BP in the quantum-confined Franz-Keldysh regime can improve maximal attainable absorption and power efficiency compared to its graphene counterpart. 
\end{abstract}
\maketitle

\emph{\textbf{Introduction---}} The mid-infrared (MIR) regime contains the fingerprints of many common molecular vibrations and covers several atmospheric transmission windows, making it important for spectroscopic molecular analysis, sensing, and free-space optical communications~\cite{Soref:1986,Stanley:2012}. As such, integrated photonic solutions that can operate between $\lambda$ = 2-10$\,\mu$m are of great technological importance. In particular, progress has been made in components such as broadband source and frequency comb with on-chip form-factor~\cite{Swiderskin:2014,Schliesser:2012}, $\mathrm{Si_{3}N_{4}}$ and SiGe-based low-loss optical waveguides, and photodetectors utilizing low-bandgap materials~\cite{Singh:2014}. However, the realization of MIR optical modulators, which require material platforms with versatile opto-electronic properties, remain challenging.

Materials with superior electro-optic properties for modulation have experienced remarkable developments in the telecommunication spectrum (0.8-1.7\,$\mathrm{\mu}$m). This can be attributed to the advents of bandgap engineering in III-V heterostructures as well as quantum-confined stark effect (QCSE), where the absorption band edge of quantum well (QW) shifts towards lower energy in the presence of transverse electric field~\cite{Miller:1984,Kuo:2006,Kuo2:2005,Arad:2003,Liu_qcse:2003}. With recent focus on utilizing the Si-on-insulator (SOI) platform to implement and integrate all possible optoelectronic functions, numerous developments have been made in using Si/SiGe heterostructures~\cite{Liu:2008,Kim:2014}. However, as these materials cannot operate beyond the near-IR, materials with suitable physical and opto-electronic properties for optical modulation on SOI is still lacking in the MIR.

Recently, interest in multilayer black phosphorus (BP) thin-film has reemerged~\cite{Li:2014,Liu:2014,Xia:2014,Koenig:2014,Morita:1986,Buscema:2014, Low:2014}. In its bulk form, BP is a semiconductor with a direct bandgap of 0.3\,eV and its measured Hall mobilities approaches 10,000\,$\mathrm{cm^{2}/Vs}$. In its thin-film form, the optical spectra of multilayer BP varies with thickness as well as light polarization across mid- to near-IR frequencies~\cite{Low2:2014,Qiao:2014,Andres:2014,Tran:2014}. Similar to graphene, the reduced dimensionality in BP allows the Pauli-blocked Burstein-Moss shift to manifest through increased doping~\cite{Low2:2014,Moss:1954,Burstein:1954,Liu:2012}. Recent electrical measurements on multilayer BP showed encouraging results~\cite{Li:2014,Liu:2014,Xia:2014,Koenig:2014,Buscema:2014}. Moreover, study on BP photo-transistor have demonstrated hyperspectral light detection covering both visible and IR frequencies~\cite{Engel:2014,Buscema:2014}. This is followed by the report of a waveguide-integrated multilayer BP photodetector that has intrinsic responsivity of 135\,mA$W^{-1}$. Moreover, it showed orders of magnitude reduction in dark current compared to its graphene-counterpart, thereby revealing BP's potential for improving the power-efficiency of optoelectronic components.  

In this work, we examine the viability of BP thin-films as electro-optic material for modulation in the MIR frequencies. Our calculations show that, under an applied out-of-plane electric field, the interplay of field-induced quantum-confined Franz-Keldysh (QCFK) effect and carrier-induced Burstein-Moss shift (BMS) leads to versatile optoelectronic attributes: BP's absorption spectra may undergo red-, blue-, or bidirectional shift depending on doping level, wavelength, and BP film thickness. Through simulation of a simple optical modulator with BP overlaid over a silicon nanowire, we elucidate BP's potential for improving maximal attainable absorption and power efficiency compared to its graphene counterpart. 

\begin{figure}[t]
	\includegraphics[width=0.48\textwidth]{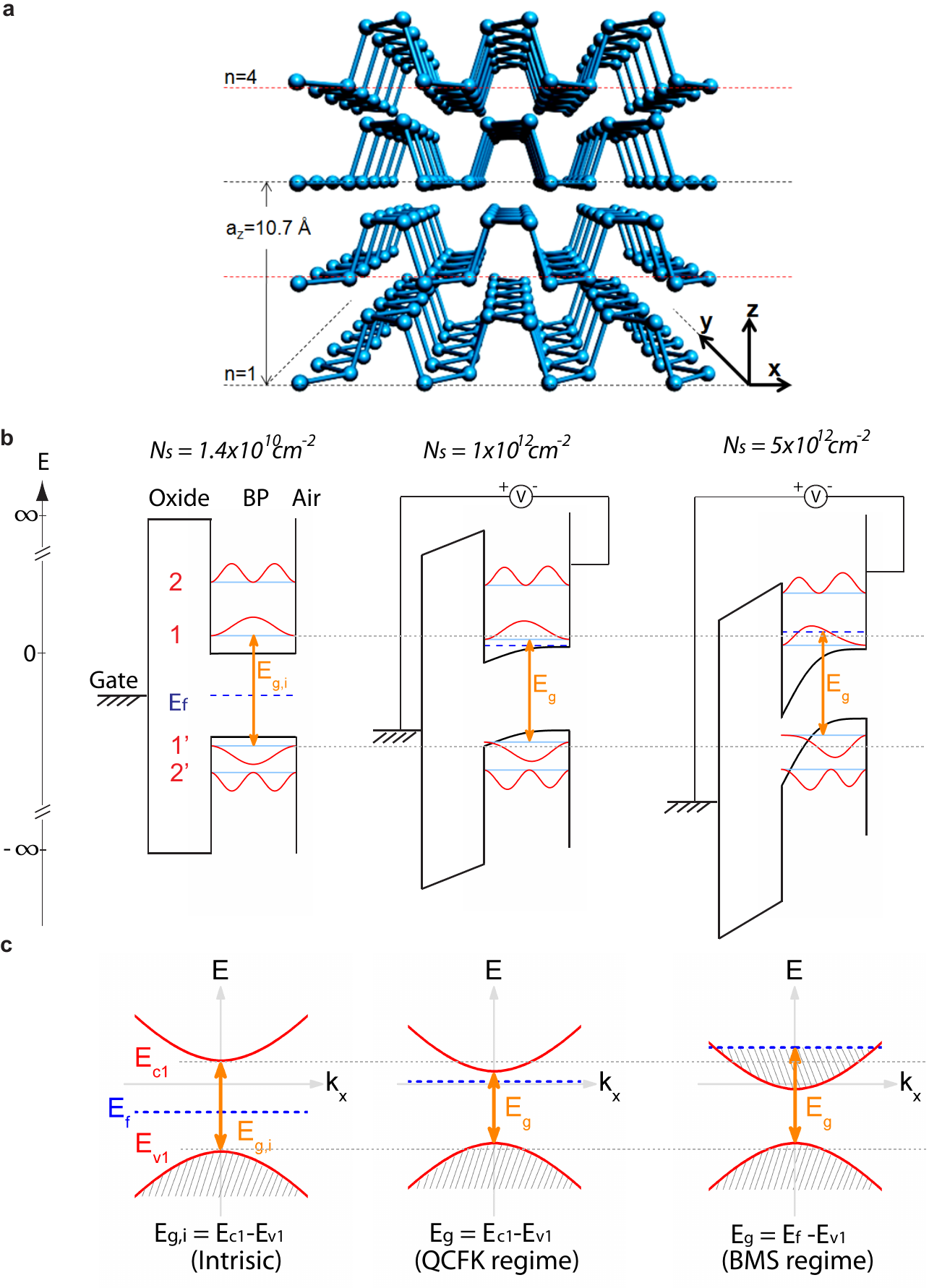}%
	\centering
	\caption{\textbf{a} Lattice structure of BP. Layer numbers \textit{n} are indicated. \textit{$a_{z}$} is the lattice constant in the out-of-plane direction. Thickness of the multilayer BP is then given by $n\times a_{z}/2$. \textbf{b} Schematic, energy band diagram, and wavefunctions of 5-nm-thick BP QW with $N_{s}$ = $1.4\times10^{10}$ (Intrinsic regime), $1\times10^{12}$ (QCFK regime), and $5\times10^{12} \mathrm{cm^{-2}}$ (BMS regime). Results are computed through self-consistent Schr$\mathrm{\ddot{o}}$dinger-Poisson calculation. The energy zero is chosen to be the at the bottom and center of the conduction band. \textbf{c} Energy dispersion diagrams for the lowest/highest conduction/valence subband under doping conditions listed above. The level of carrier occupancy is indicated by the shaded region. The corresponding formula for calculating $E_{g}$ are also listed.}%
\end{figure}

\emph{\textbf{BP quantum well electrostatics---}} BP has an orthorhombic crystal structure consisting of puckered layers as illustrated in Fig.\,1(a). It exhibits anisotropic in-plane optical properties, with the armchair direction ($x$) corresponding to the lower in-plane effective mass direction~\cite{Morita:1986}. Moreover, it has a direct gap at $\Gamma$ point, estimated to be between 2 and 0.3\,eV for monolayer and bulk BP respectively~ \cite{Morita:1986,Rudenko:2014,Tran:2014}. Cyclotron resonance experiments on bulk BP~\cite{Narita:1983} have found the out-of-plane effective masses to be considerably smaller than that of TMDs~\cite{Mattheiss:1973} and graphite~\cite{Wallace:1947}. Adopting an average of experimental~\cite{Narita:1983} and theoretical~\cite{Low2:2014} values, the electron and hole out-of-plane masses are about $m_{cz}$ = 0.2 $m_{0}$ and $m_{vz}$ = 0.4 $m_{0}$ respectively. 

Here, we examine the electrostatics in a 5-nm BP QW under the presence of an out-of-plane electric field. We describe its electronic structure around the $\Gamma$ point within the k$\cdot$p theory two-band Hamiltonian described elsewhere \cite{Low2:2014,Rodin:2014}. See Methods for detailed description. With an infinite potential barrier assumption, we solved for the QW electrostatics self-consistently~\cite{Stern:1970}. In the intrinsic case, i.e. when the Fermi energy ($E_{f}$) is approximately located at midgap, the electron carrier density ($N_{s}$) of the 5-nm BP QW is calculated to be $1.4\times10^{10} \mathrm{cm^{-2}}$, albeit compensated by the same amount of hole density. Here, the intrinsic case also corresponds to flatband condition. In this work, we define the optical bandgap, $E_{g}$, to be the optical transition energy between the highest filled state in the valence band and the lowest unoccupied state in conduction band. For our intrinsic 5\,nm BP film, this is determined to be $E_{g,i}$ = $E_{c1}-E_{v1}$ = 0.62\,eV, where $E_{c1}$ and $E_{v1}$ are the first conduction and valence subbands respectively. 

\begin{figure}
	\includegraphics[width=0.48\textwidth]{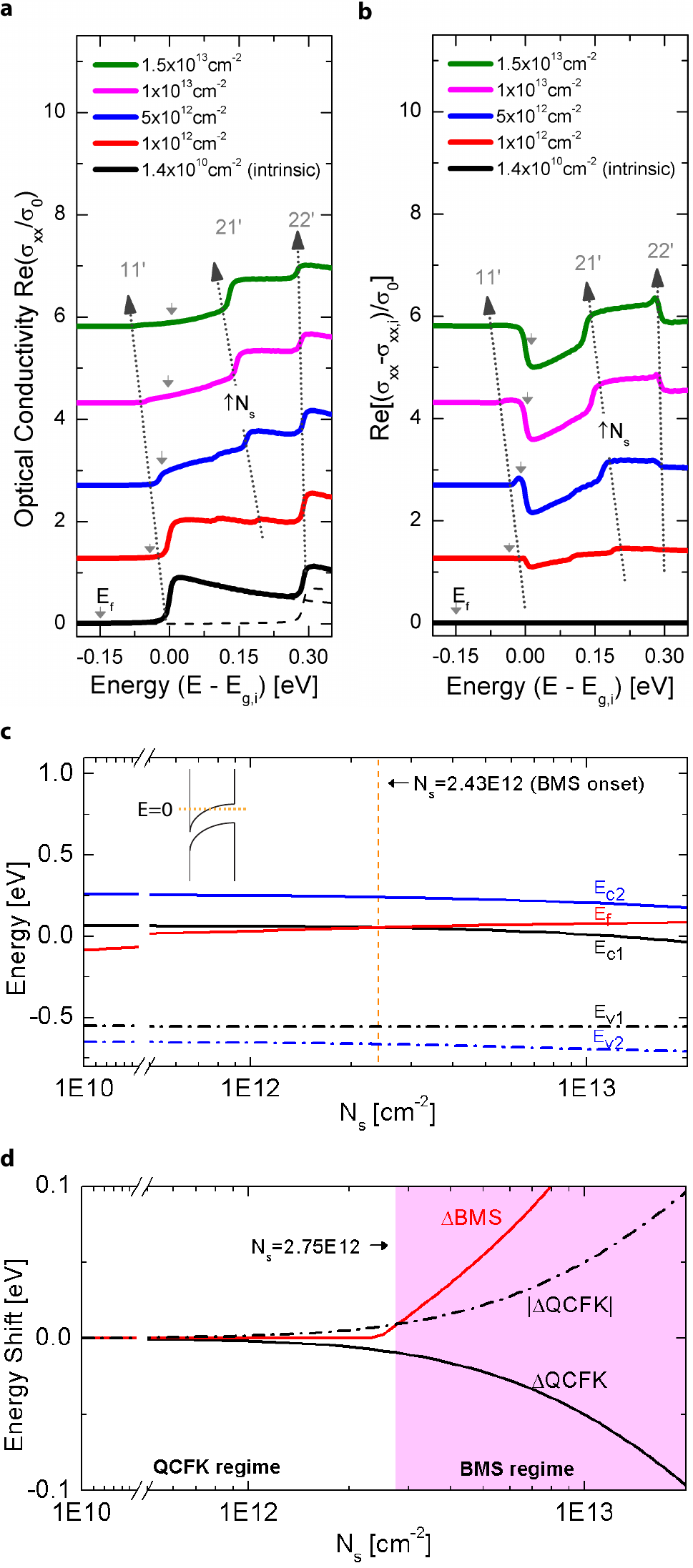}%
	\centering
	\caption{ \textbf{a} Evolution of $\sigma_{xx}$ of a 5-nm BP QW due to increasing $N_{s}$ level. Conductivities and photon energies are normalized with respect to $\sigma_{0}=e^{2}/4h$ and $E_{g,i}$ (0.62 eV) respectively. The intersubband contributions to $\sigma_{xx}$ are illustrated for intrinsic BP ($N_{s} = 1.4\times10^{10} cm^{-2}$). \textbf{b} Differential conductivity spectra, ($\sigma_{xx}$-$\sigma_{xx,i}$)/$\sigma_{0}$, of the 5-nm BP QW. \textbf{c,d} Calculated QW energies and shifts in $E_{g}$ due to QCFK effect and BMS (described by eq. (2) and (3) respectively).  }%
\end{figure}

Fig.\,1(b) presents the energy band diagram of the 5\,nm BP QW in the out-of-plane direction using armchair effective mass. With the application of a positive external gate bias, the out-of-plane electric field leads to band-bending across the QW, bringing the electron and hole subbands closer in energy, hence effectively reducing $E_{g}$ ~\cite{Miller:1984}. Such bandgap shrinkage due to Stark shift have been recently observed experimentally~\cite{Kim:2015} and can be described by QCFK effect~\cite{Miller:1986} if excitonic effect is not considered. Concurrently, the induced $N_{s}$ leads to the formation of 2D electron gas at the BP-oxide interface and also raises the $E_{f}$ of the QW. As the electron gas becomes more degenerate, Pauli blocking of optical transitions lead to broadening of $E_{g}$, as described by BMS~\cite{Low:2014,Wang:2008,Li:2008}. The former mechanism leads to a red-shift, while the latter effect appears as a blue-shift.  The dispersion diagrams shown in Fig.\,1(c) illustrates these optical processes. The net shift in $E_{g}$ due to an external bias can therefore be described by
\begin{equation}
\Delta E_{g} \propto \Delta_{QCFK} + \Delta_{BMS}.   
\end{equation}
The bias-induced band-bending has a negative contribution to $E_{g}$ and can be approximated by
\begin{equation}
\Delta_{QCFK} \sim \delta(E_{c1}-E_{v1}).
\end{equation}
The shift in $E_{g}$ in a degenerate electron gas due to Pauli blocking, ignoring the conduction and valence bands asymmetry, can be approximated by
\begin{equation}
\Delta_{BMS} \sim 2\delta(E_{f} - E_{c1})U(E_{f} - E_{c1})
\end{equation}
for electron-doped BP, where $U(...)$ denotes the heaviside step function.
When $E_{f}$ raises beyond $E_{c1}$, $\Delta_{BMS}$ acquires a positive contribution to $E_{g}$. A similar expression holds for the hole-doped case. 
\begin{figure*}
	\includegraphics[width=1.0\textwidth]{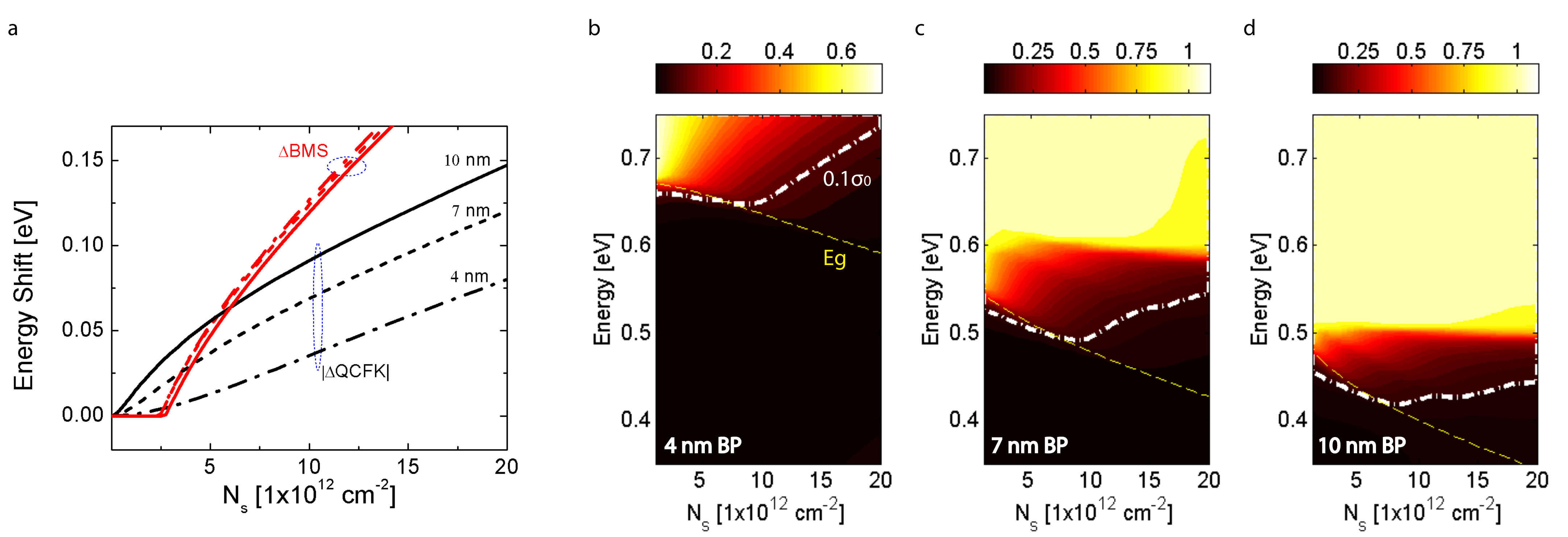}%
	\caption{ \textbf{a} Calculated shift in $E_{g}$ due to $\Delta_{BMS}$ and $\Delta_{QCFK}$ for 4-nm, 7-nm, and 10-nm BP QWs, plotted as a function of $N_{s}$. \textbf{b-d} Optical conductivities ($\sigma_{xx}$) of the respective QWs, plotted as a function of $N_{s}$ and photon energy (\textit{E}). Conductivities are normalized with respect to $\sigma_{0}=e^{2}/4h$.}%
\end{figure*}

\emph{\textbf{Optical conductivity---}} 
We calculate BP's optical conductivity tensor ($\sigma_{ij}$) using the Kubo formula within an effective low-energy Hamiltonian as described elsewhere~\cite{Low2:2014}. We consider a range of positive bias, up to carrier concentration of $N_s\sim 1.5\times10^{13}\,$cm$^{-2}$, a value routinely obtained in experiments with layered materials. Temperature is taken to be 300\,K in all calculations, and a phenomenological broadening term of 5\,meV is assumed to account for finite electron lifetimes. 
Fig.\,2(a) studies the evolution of the $\sigma_{xx}$ of a 5-nm BP QW with different $N_s$, where $x$ is the crystal axes with higher optical conductivity. The corresponding differential conductivity spectra ($\sigma_{xx}$-$\sigma_{xx,i}$)/$\sigma_{0}$ are shown in Fig. 2(b). We normalized $\sigma_{xx}$ with respect to $\sigma_{0} = e^{2}/4\hbar$, i.e. the universal conductivity of graphene~\cite{Nair:2008,Mak:2008}. 
The photon energies $E$ is displayed with respect to $E_{g,i}$. BP's $\sigma_{xx}$ not only shows strong doping dependence similar to that of graphene (see accompanying Suppl. Info), but also exhibits oscillatory behaviour, which can be traced to the underlying electronic subbands structure. We explain these trends in more detail below, by examining optical transitions between electron and hole subbands across the band gap i.e. $E_{cn}$ and $E_{vm}$:

$\mathbf{E_{c1}-E_{v1}~transition:}$ For intrinsic BP, the oscillator strength is non-zero only for optical transitions between subbands of the same indices i.e. $m=n$. Hence, the lowest energy optical transition, $E_{c1}-E_{v1}$, defines the optical bandgap $E_g$, as apparent in Fig.\,2(a). With increased $N_{s}$, the red-shift of the optical spectra is clearly observed. The energy levels of the QW as a function of $N_{s}$ is displayed Fig.\,2(c). The energy zero is chosen to be at the center of  the conduction band bottom (see inset). The narrowing of $E_{g}$ as described by QCFK effect is attributed to the effective lowering of $E_{c1}$ and $E_{v1}$. For progressively higher $N_{s}$, hence stronger external field across the QW, the electron and hole wavefunctions start to shift to opposite sides of the QW (Fig.\,1(b))~\cite{Miller:1984}. Consequently, the transition starts to become quenched and its optical band edge becomes smeared, reflected also in Fig. 2(a).

$\mathbf{E_{c2}-E_{v1}~transition:}$ With sufficient electric field strength, the wavefunction overlap between $m\neq n$ subbands becomes finite, resulting in the appearance of added "ripples" in $\sigma_{xx}$ (Fig. 2(a)). The induced tail and the additional oscillatory features below and above $E_{g,i}$ are both characteristics of the QCFK effect \cite{Miller:1986}. For $N_{s} > 5\times10^{12} \mathrm{cm^{-2}}$, the $E_{c2}-E_{v1}$ transition becomes the dominant feature in the optical spectra. The field-induced red-shift similar to that of $E_{c1}-E_{v1}$ transition is observed, but the absorption edge becomes more abrupt with increased doping, thereby elucidating the contribution from BMS. 

$\mathbf{E_{c2}-E_{v2}~transition:}$ For this 5-nm BP film, the second subband energies are larger than the band bending of the QW. Moreover, only the $E_{c1}$ subband is populated for the chosen doping range. Thus, the position of this transition remains almost invariant (Fig. 2(a)).

Figure 2(d) studies the doping-dependence of $\Delta_{QCFK}$ and $\Delta_{BMS}$ in the 5-nm BP QW. Depending on the doping level, the change in $E_{g}$ may be categorized into two regimes, each dominated by different electro-absorption mechanisms. For low $N_{s}$, the $E_{f}$ level is lower than $E_{c1}$, indicating a red-shift of $E_{g}$ as only $\Delta_{QCFK}$ provides finite contribution. At $N_{s} = 2.43\times10^{12} \mathrm{cm^{-2}}$, the $E_{f}$ and $E_{c1}$ levels crosses, after which $\Delta_{BMS}$ increases rapidly, eventually matches and exceeds $|\Delta_{QCFK}|$ at $N_{s} = 2.75\times10^{12} cm^{-2}$. For high $N_{s}$, blue-shift of $E_{g}$ starts to manifest, albeit counteracted by $\Delta_{QCFK}$. The evolution and transition between the two absorption regimes are further elucidated in the differential conductivity spectra (Fig. 2(b)). For $N_{s} = 1.5\times10^{12} \mathrm{cm^{-2}}$, the QCFK-induced red-shift in $E_{g}$ results in an asymmetric line shape, where an increase/decrease of $\Delta\sigma_{xx}$ at photon energies below/above $E_{g,i}$ can be observed. Conversely, for $N_{s}= 1\times10^{13} \mathrm{cm^{-2}}$, a large decrease of $\Delta\sigma_{xx}$ above $E_{g,i}$, a characteristic of BMS, is obtained.

Other than doping, the optical absorption spectra of BP thin films also vary sensitively with the number of layers. Figure 3(a) shows the change in $E_{g}$ as a function of doping, for BP QW with thickness of 4, 7, and 10 nm. The difference between $\Delta_{BMS}$ and $|\Delta_{QCFK}|$ for high $N_{s}$ level reduces with increased thickness as wider QWs have lower subband energies and thus are more strongly influenced by external electric field. The $\sigma_{xx}$ spectra for the QWs are displayed from Fig. 3(b) to 3(d), where $E_{g}$ and optical band edge, defined as $0.1 \times\sigma_{0}$, are also indicated. In general, the magnitude of $\sigma_{xx}$ increases with BP thickness. For low $N_{s}$, the optical band edge shifts towards lower energy and follows closely with $E{g}$. The shift is more abrupt for thicker QW due to stronger QCFK effect. For higher $N_{s}$, the band edge deviates from $E_{g}$ and shift towards higher energy due to BMS. The shift is more abrupt for thinner film due to more rapid band filling, since the subbands are more apart in energy.

Note that for BP films thinner than those studied in this paper, bandgap renormalization and strong excitonic effects are also expected due to reduced screening~\cite{Tran:2014, Chaves:2015,Wang:2015}. This may lead to further band edge shift and enhanced light absorption not accounted for in present calculations. 

\emph{\textbf{BP as an active optical layer for modulation---}} For proof-of-concept, we study the performance of a simple modulator design, consisting of a BP thin-film overlaid on top of a Si nanowire, separated by a 7 nm $\mathrm{Al_{2}O_{3}}$ spacer~(Fig 4(a)). To maximize the overlap between the optical mode and the actively-modulated BP layer, the transverse-magnetic mode (Fig. 4(a) inset) is chosen for modulation and BP's crystal axes in the amrchair direction is assumed to align with the direction of light propagation. Such modulator configuration was reported in the initial demonstration of graphene-assisted modulator~\cite{Liu:2012}. 
\begin{figure*}[t!]
	\includegraphics[width=1.0\textwidth]{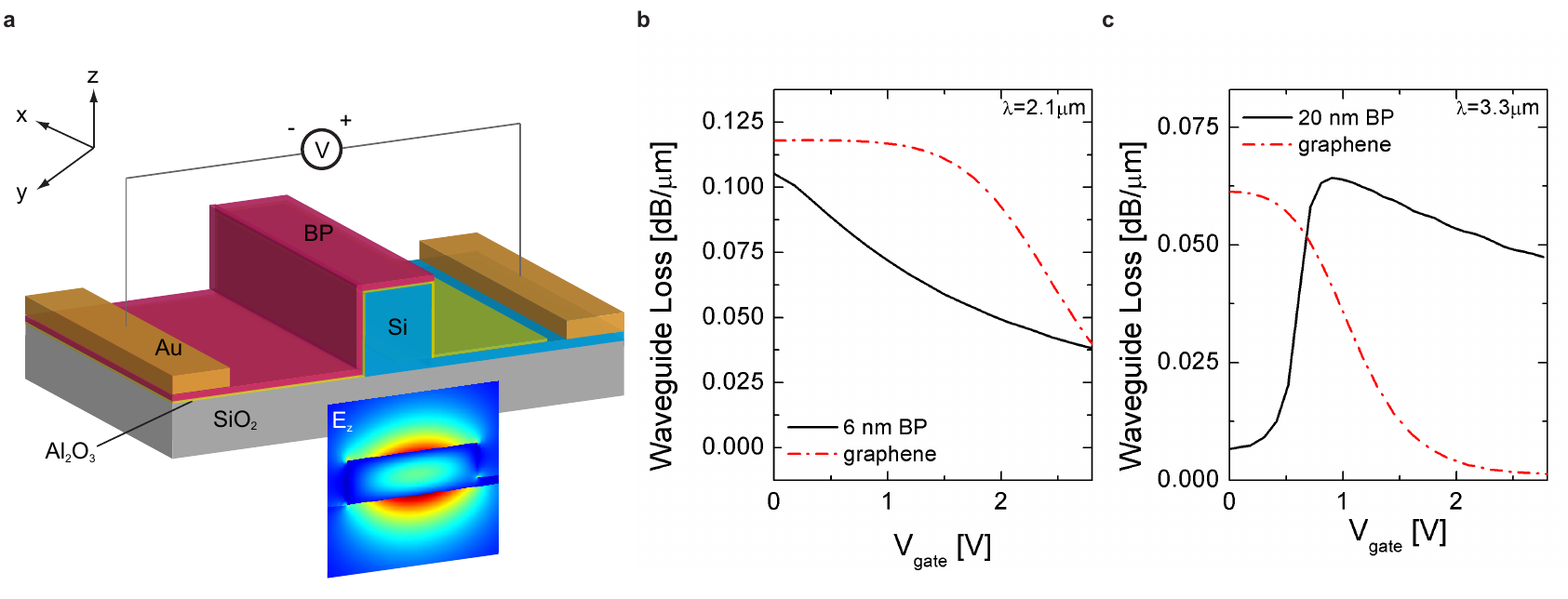}%
	\centering
	\caption{ \textbf{a} Schematic of BP-assisted, travelling-wave electro-absorption modulator. The Si waveguide is connected to a partially-etched, 50-nm-thick Si layer that serves as a contact electrode. The Si layer is assumed to be undoped. The effective numerical thickness of BP is taken to be 0.7 nm~\cite{Liu:2012}. The transverse-magnetic mode profile is shown in inset. \textbf{b} Waveguide loss for modulator integrated with 6-nm BP active layer, operating at $\lambda = 2.1 \mu$m. The thickness and width of the Si core are 350 nm and 850 nm respectively. The performance of the modulator where BP is replaced with monolayer graphene is displayed for comparison. \textbf{c} Waveguide loss for modulator integrated with 20-nm BP active layer, operating at $\lambda = 3.3 \mu$m. The thickness and width of the Si core are 510 nm and 1200 nm respectively.    
	}%
\end{figure*}

Figure 4(b) shows the static electro-optical modulator response using BMS-dominant, 6-nm BP thin film, operating at $\lambda = 2.1 \mu m$. For comparison, the modulator response with BP layer replaced with monolayer graphene is displayed. The optical properties of graphene is calculated via Kubo's formula ~\cite{Falkovsky:2007}, with phenomenological broadening of 5 meV (Suppl. Fig. 1). While the waveguide loss associated with 6-nm BP is more sensitive to bias at lower gate voltage ($V_{gate}$), graphene enables more rapid decay at higher $V_{gate}$. This is a consequence of the quenching of the BMS effect due to the QCFK mechanism. 

Conversely, by increasing BP layer thickness, the $V_{gate}$ required to induce the on-set of BMS increases, thereby allowing stronger QCFK effect to manifest. Figure 4(c) shows the response of a 20-nm BP-assisted modulator, operating at $\lambda = 3.3 \mu m$. Recent experiment have shown low-loss light propagation through SOI platform near 3 $\mathrm{\mu m}$, showing promise for integrated MIR data communication~\cite{Mashanovich:2011}. Here, the waveguide loss increases and decreases for gate bias ($V_{gate}$) below and above 0.89 V respectively, indicating a transition between QCFK and BMS regimes. The change in waveguide loss is more rapid in the QCFK regime, since the modulation is via an effective change in transition energy gap, below which there are no available electronic density of states for the optical transition. Conversely, BMS relies on Pauli blocking, where the optical transition edge is smeared in energy by kT. Operating in the QCFK regime, the performance of 20-nm BP compares favorably against graphene: (1) The maximal attainable absorption is 0.65 and 0.61 dB/$\mu$m for BP and graphene respectively. (2) Defining the modulator power consumption ($P$) as the voltage swing required to reduce the waveguide loss from maximally lossy state (OFF-state) to 0.01 dB/$\mu$m (ON-state), BP-assisted modulator only requires $P$=0.59 V. This is a 62 \% reduction compared to the use of BMS effect in its graphene counterpart ($P$=1.56 V). 

One popular figure-of-merit (FOM) for optical modulators is the ratio of extinction ratio over insertion loss, where extinction ratio is the difference between the ON-state and the OFF-state losses while insertion loss is the ON-state loss~\cite{Schaevitz:2012}. Although QCFK is quenched by BMS in the current modulator design, the device achieves a FOM of 5.5 and uses only a single BP QW that couple evanescently with the optical mode. This exceeds the FOM = 1.6-4 as obtained with QCSE-based SiGe modulators~\cite{Schaevitz:2012}, which places multiple ($\textgreater$10) QWs in a carrier-depleted PIN junction as a crystalline, multi-layer waveguide core. Thus, layering of BP with other 2D materials to engineer analogous heterojunctions may further enhance the performance of BP-assisted modulators. 

\emph{\textbf{Conclusion---}} In summary, we demonstrated the versatility of BP's electro-absorption characteristics in the MIR. Controlled by the interplay of QCFK and BMS effects, the optical bandgap may undergo blue-shift, red-shift, or bidirectional-shift for a given doping range, film thickness, and wavelength. This is afforded by the reduced dimensionality and finite bandgap of BP, which departs from the unidirectional shift observed in graphene and traditional III-V semiconductors. Simulation of a simple modulator design indicates that QCFK effect is superior than BMS as an electro-absorption mechanism in BP thin-films. Operating in QCFK regime leads to enhanced maximal attainable absorption and reduced power consumption compared to its graphene counterpart. Overall, the gate-tunable optical characteristics make multilayer BP an attractive, alternative material platform for integrated optoelectronic systems in the MIR.

\emph{\textbf{Methods---}} \textit{Optical conductivity model of multilayer black phosphorus:} Based on k $\cdot$ p theory and symmetry arguments, the in-plane, low-energy Hamiltonian around the $\Gamma$ point can be described as \cite{Rodin:2014}:
\begin{equation}
	H =
	\begin{pmatrix}
		E_{c} + \eta_{c}k_{x}^{2} + \nu_{c}k_{y}^{2} & \gamma k_{x} \\
		\gamma k_{x} & E_{v} - \eta_{v}k_{x}^{2} - \nu_{v}k_{y}^{2} \\
	\end{pmatrix}
\end{equation}
where $E_{c}$ and $E_{v}$ are the energies of the conduction and valence band edges, while $\gamma$ describes the effective couplings between the two bands. The $\eta_{c,v}$ and $\nu_{c,v}$ terms are related to the in-plane effective masses, given by $m_{cx} = \hbar^{2}/[2\gamma^{2}/(E_{c}-E_{v})+\eta_{c}]$ and $m_{cy}=\hbar^{2}/2\nu_{c}$ for electrons. They are chosen such that they yield the known effective masses in the bulk limit ($m_{cx}$=$m_{vx}$=0.08$m_{0}$, $m_{cy}$=0.7$m_{0}$, and $m_{vy}$=1.0$m_{0}$)~\cite{Morita:1986,Narita:1983} and monolayer BP ($m_{cx}$=$m_{vx}\approx$0.15$m_{0}$)\cite{Rodin:2014}. Analogous expression applies for the hole case. Using known anisotropic effective masses for monolayer and bulk BP films \cite{Morita:1986,Low:2014,Narita:1983}, we set: $\eta_{c,v}$ = $\hbar^{2}/0.4m_{0}$, $\nu_{c}$ = $\hbar^{2}/1.4m_{0}$, $\nu_{v}$ = $\hbar^{2}/2.0m_{0}$, and $\gamma$ = $4a/\pi$ eVm. The value of $\beta$ is taken to be $2a^{2}/\pi^{2}$ eV/$m^{2}$, where a = 2.23 \AA \hspace{0.05cm} and $\pi/a$ is the width of the BZ in the \textit{x} direction.
\paragraph{}
Due to quantum-confinement, $E_{c,v}$ in Eq. (4) needs to be replaced with $E_{c,v}^{j}$, where \textit{j} denotes the subband number. Moreover, additional confinement energies are incorporated in our model to reproduce the predicted energy gap of the BP film, of 2 and 0.3 eV in the monolayer and bulk limit respectively \cite{Low:2014,Tran:2014}. For electrons, they are given by $\delta E_{c}^{j} = j^{2}\hbar^{2}\pi^{2}/2m_{cz}t_{z}$, where $t_{z}$ is the thickness of the BP film, and $m_{cz}$ = 0.2 $m_{0}$ is the electron effective mass along \textit{z} direction \cite{Low:2014,Narita:1983}. Analogous expression applies for the hole case ($m_{vz}$ = 0.4 $m_{0}$). 
\paragraph{}
The subband energies can be obtained through self-consistent solution of Schr$\mathrm{\ddot{o}}$dinger-Poisson equations~\cite{Stern:1970}:  
\begin{equation}
	\begin{aligned}
	\frac{d^{2}\varphi(z)}{dz^{2}}= \frac{-4\pi e^{2}}{\epsilon_{BP}}\times \qquad\qquad\qquad\qquad\qquad\qquad\qquad\qquad\\ 
	\sum\limits_{j} \Big[\frac{m_{DOS}^{j}k_{B}T}{\pi\hbar^{2}}\mbox{ln}\Big(1+\mbox{exp}(\frac{E_{F}-E_{c}^{j}}{k_{B}T})\Big)|\psi(z)^{2}|\Big]
	%
	\end{aligned}
\end{equation}
where $E_{F}$ is the Fermi level, $m_{DOS}^{j}$ = $\sqrt{(m_{cx}^{j}m_{cy}^{j})}$ is the subband density-of-states mass, and $\epsilon_{BP}$ = 8.3 is the out-of-plane dielectric constant of BP~\cite{Morita:1986}.
\paragraph{}
Finally, the optical conductivity of BP can be evaluated via the Kubo formula:
\begin{equation}
	\begin{split}
		\sigma_{\alpha\beta}(\textbf{q},\omega) = & -i \frac{g_{s}he^{2}}{(2\pi)^{2}}\sum\limits_{ss{'}jj^{'}}\int d\textbf{k}\frac{f({E_{sj\textbf{k}}})-f({E_{s^{'}j^{'}\textbf{k}^{'}}})}{E_{sj\textbf{k}}-E_{s^{'}j^{'}\textbf{k}^{'}}} \\
		& \times \frac{\langle \Phi_{sj\textbf{k}}|\hat{\nu}_{\alpha}|\Phi_{s^{'}j^{'}\textbf{k}^{'}} \rangle
			\langle \Phi_{s^{'}j^{'}\textbf{k}^{'}}|\hat{\nu}_{\beta}| \Phi_{sj\textbf{k}} \rangle
		}{E_{sj\textbf{k}}-E_{s^{'}j^{'}\textbf{k}^{'}}+\hbar\omega+i\eta}
	\end{split}
\end{equation}
where $\hat{\nu}_{\alpha}$ is the velocity operator defined as $\hbar^{-1}\delta_{k\alpha}H$, $g_{s}$ = 2 accounts for the spin degeneracy, and $\eta$ is a phenomenological broadening term to account for finite damping, assigned to be 5 meV. $E_{sjk}$ and $\Phi_{sjk}$ are the eigenenergies and eigenfunctions of \textit{H}. $f$(···) is the Fermi-Dirac distribution function. The indices \{\textit{s,s}\}=$\pm$1 denote conduction/valence band. 
\paragraph{}The dielectric constant of BP ($\epsilon$) can subsequently be calculated via:
\begin{equation}
\epsilon = 1 - \frac{\sigma}{j\omega\epsilon_{0}\Delta}
\end{equation}
where $\epsilon_{0}$ is the vacuum permittivity and $\Delta$ = 0.7 nm is used as the numerical effective thickness of BP~\cite{Liu:2012}.

\emph{\textbf{Supplemental Information---}} Details of graphene's optical model and the evolution of its conductivity spectrum as a function of electron carrier density.

\end{document}